\documentclass[num-refs]{nbdt-article}

\newcommand{\argmin}{\operatornamewithlimits{argmin}}
\pdfoutput=1

\papertype{Original Article}
\paperfield{Theory}

\title{What does the free energy principle tell us about the brain?}

\author[1\authfn{1}]{Samuel J. Gershman}

\affil[1]{Department of Psychology and Center for Brain Science, Harvard University, Cambridge, MA, 02138, USA}

\corraddress{Northwest Laboratories, 52 Oxford St., Room 295.05, Cambridge, MA, 02138, USA}
\corremail{gershman@fas.harvard.edu}

\fundinginfo{This work was supported by a research fellowship from the Alfred P. Sloan Foundation.}

\runningauthor{Samuel J. Gershman}

\begin{document}

\maketitle

\begin{abstract}
The free energy principle has been proposed as a unifying account of brain function. It is closely related, and in some cases subsumes, earlier unifying ideas such as Bayesian inference, predictive coding, and active learning. This article clarifies these connections, teasing apart distinctive and shared predictions.

\end{abstract}

\pdfoutput=1

\section{Introduction}

The free energy principle (FEP) states, in a nutshell, that the brain seeks to minimize surprise \cite{friston10}. It is arguably the most ambitious theory of the brain available today, claiming to subsume many other important ideas, such as predictive coding, efficient coding, Bayesian inference, and optimal control theory. However, it is precisely this generality that raises a concern: what exactly does FEP predict, and what does it not predict? Addressing this concern is not easy, because the assumptions underlying applications of FEP are malleable (e.g., different applications use different generative models, different algorithmic approximations, and different neural implementations). Moreover, some of these assumptions are shared with other theories, and some are idiosyncratic; some assumptions are central to the theory, and others are \emph{ad hoc} or made for analytical convenience.

This article systematically deconstructs the assumptions underlying FEP, with the goal of identifying what its distinctive theoretical claims are. As will become clear, FEP does not have a fixed set of distinctive claims. Rather, it makes different claims under different sets of assumptions. This is not necessarily a bad thing, provided we can verify these assumptions in any particular application and thus render the theoretical assumptions falsifiable.

Before proceeding, we must address two qualms with this deconstructive approach. Some proponents of FEP might reasonably argue that identifying distinctive theoretical claims is pointless; the whole point of a unifying theory is to unify claims, not distinguish them. However, the fundamental issue here is not whether one theory is better than another, but how to assign credit and blame to different theoretical assumptions. If FEP fails to account for the data, is that attributable to the assumption that the brain is Bayesian, or a particular algorithmic implementation of Bayesian inference, or particular assumptions about the probabilistic model? Only by answering such questions can we understand the successes and failures of a unifying theory, devise suitable tests of its assumptions, and identify ways to improve the theory.

Another qualm with this approach is based on the argument that FEP is not a theory at all, in the sense that a theory constitutes a set of falsifiable claims about empirical phenomena. What makes it a \emph{principle}, rather than a theory, is that it constitutes a set of self-consistent statements in a formal mathematical system. In this sense, a principle cannot be falsified through the study of empirical phenomena. A theory designates correspondences between formal statements and empirical phenomena, and thus can be falsified if the theory makes incorrect predictions on the basis of these correspondences. Viewed in this way, FEP is unobjectionable: its mathematical soundness is sufficient demonstration of its credentials as a principle. Here we will be concerned with its credentials as a theory, and therefore we will pay particular attention to specific implementations (process models).

\section{The Bayesian brain hypothesis}

As a prelude to FEP, it will be helpful to briefly describe the Bayesian brain hypothesis \cite{lee03,knill04,doya07}, which can be expressed in terms that are more familiar to neuroscientists, and is in fact equivalent to FEP under certain conditions (as elaborated in the next section). The first claim of the Bayesian brain hypothesis is that the brain is equipped with an internal (or ``generative'') model of the environment, which specifies a ``recipe'' for generating sensory observations (denoted by $o$) from hidden states (denoted by $s$). This internal model may not be represented explicitly anywhere in the brain; the claim is that the brain computes ``as if'' it had an internal model. In order for the Bayesian brain hypothesis to have any predictive power, it is necessary to make specific assumptions about the structure of the internal model.

There are two components of the internal model that need to be specified. First, hidden variables are drawn from a \emph{prior distribution}, $p(s)$. For example, the hidden state might be the orientation of a line segment on the surface of an object, and the prior might be a distribution that favors cardinal over oblique orientations \cite{girshick11}. Second, the sensory observations are drawn from an observation distribution conditional on the hidden state, $p(o|s)$. For example, the hidden line orientation is projected onto the retina and then encoded by the firing of retinal ganglion cells. This encoding process might be noisy (due to stochasticity of neural firing) or ambiguous (due to the optical projection of three dimensions onto the two-dimensional retinal image), such that different settings of the hidden state could plausibly ``explain'' the observations to varying degrees. These degrees of plausibility are quantified by the \emph{likelihood}, the probability of the observations under the observation distribution given a hypothetical setting of the hidden state.

The second claim of the Bayesian brain hypothesis is that the the prior and the likelihood are combined to infer the hidden state given the observations, as stipulated by Bayes' rule:
\begin{align}
p(s|o) = \frac{p(o|s) p(s)}{p(o)},
\end{align}
where $p(s|o)$ is the \emph{posterior distribution} and $p(o) = \sum_{s} p(o|s) p(s)$ is the \emph{marginal likelihood} (for continuous states, the summation is replaced with integration). We can think of Bayes' rule as ``inverting'' the internal model to compute a belief about the hidden state of the environment given the observations.

The Bayesian brain hypothesis can be naturally extended to settings where an agent can influence its observations by taking actions according to a policy $\pi$, which is a mapping from observations to a distribution over actions. In the simplest variant, an agent chooses a policy that maximizes \emph{information gain}:
\begin{align}
\mathcal{I}(\pi) =  \sum_{o} p(o|\pi) \mathcal{D}[p(s|o,\pi)||p(s|\pi)],
\end{align}
where $o$ now denotes a future observation, and $D$ denotes the Kullback-Leibler (KL) divergence (also known as \emph{relative entropy}):
\begin{align}
\mathcal{D}[p(s|o,\pi)||p(s|o)] = \sum_{s} p(s|o,\pi) \log \frac{p(s|o,\pi)}{p(s|\pi)}.
\end{align}
The expression for $\mathcal{I}(\pi)$ is equivalent to ``Bayesian surprise'' \cite{itti09}, and to the mutual information between $s$ and $o$ conditional on $\pi$ \cite{cover91,lindley56}. Information maximization has been studied extensively in the cognitive psychology literature \cite{oaksford94,nelson05,tsividis14}. More generally, information maximization can be understood as a form of \emph{active learning} that has been studied extensively in the machine learning and statistics literature \cite{settles12}.

Information gain maximization is a special case of Bayesian decision theory, where the utility $u(o) = \mathcal{D}[p(s|o,\pi)||p(s|\pi)]$ of an observation corresponds to information gain. If the observations are valenced (rewards or punishments), then utilities may reflect their goodness to the agent, who seeks to maximize the expected utility:
\begin{align}
\mathbb{E}[u(o)|\pi] = \sum_o p(o|\pi) u(o).
\end{align}
This analysis can be generalized to sequential decision problems \cite[see][]{dayan08}, where an agent's actions and observations unfold over time. Typically, the goal in sequential decision problems is to maximize discounted cumulative utility (return):
\begin{align}
R(\mathbf{o}) = u(o_1) + \gamma u(o_2) + \gamma^2 u(o_3) + \cdots
\end{align}
where we have introduced a subscript denoting time-step and the bold notation $\mathbf{o} = [o_1, o_2,\ldots]$ denotes the time-series of observations. The discount factor $\gamma$ down-weights future utility exponentially as a function of temporal distance. The expected return under the posterior is then defined analogously to expected utility:
\begin{align}
\mathbb{E}[R(\mathbf{o})|\pi] = \sum_\mathbf{o} p(\mathbf{o}|\pi) R(\mathbf{o}).
\end{align}
In sequential decision problems, an agent needs to trade off gathering information to reduce uncertainty (exploration) and taking actions that yield immediate reward (exploitation). This means that preferences for information will arise instrumentally in the sequential decision setting; they need not be built explicitly into the utility function.

There are several points worth noting here before moving on:
\begin{itemize}
\item Although the Bayesian brain hypothesis has received considerable support, there are numerous empirical deviations from its claims (e.g., \cite{soltani16,rahnev18},) some of which may be rationalized by considering approximate inference algorithms \cite{gershman15}. The variational algorithms we consider below are examples of such approximations. We will not evaluate the empirical validity of the (approximate) Bayesian brain hypothesis, focusing instead on more conceptual issues related to the free energy principle.

\item The Bayesian brain hypothesis does not make any specific claims about the priors and likelihoods of an individual. Rather, the central claim concerns consistency of beliefs: a Bayesian agent will convert prior beliefs into posterior beliefs in accordance with Bayes' rule.

\item The Bayesian brain hypothesis abstracts away from any particular algorithmic or neural claims: it is purely a ``computational-level'' hypothesis. All algorithms that compute the posterior exactly give equivalent predictions with regard to the central claims of the Bayesian brain hypothesis, and likewise any neural implementation will give equivalent predictions. These equivalences do not hold, however, when we consider \emph{approximate} inference schemes, which may systematically deviate from the Bayesian ideal. We will return to this point below.
\end{itemize}

\section{The unrestricted free energy principle is Bayesian inference}

The basic idea of the FEP is to convert Bayesian inference into an optimization problem (see \cite{bogacz17} for a tutorial introduction). This idea was first developed in physics, and later in machine learning, to handle computationally intractable inference problems. The key algorithmic trick, as we will see, is to restrict the optimization problem in such a way that it is not searching over all possible posterior distributions.

Assume we have available a family of distributions $\mathcal{Q}$ (discussed further in the next section), and we can choose one distribution $q \in \mathcal{Q}$ to approximate $p(s|o)$. This leads to the following ``variational'' optimization problem:
\begin{align}
q^\ast(s) = \argmin_{q(s)} \mathcal{D}[q(s)||p(s|o)].
\end{align}
The KL divergence is 0 when $q(s) = p(s|o)$. Thus, if $p(s|o)$ is contained in the variational family $\mathcal{Q}$, then the solution of the optimization problem yields the exact posterior: $q^\ast(s) = p(s|o)$. This holds true when the variational family is unrestricted (i.e., contains all possible distributions with support on the hypothesis space).

Algorithmically, this optimization problem is not very practical because to compute the KL divergence we need access to $q(s)$---precisely the problem we are trying to solve! However, it turns out that one can reformulate this problem in a way that is more practical, based on the following identity:
\begin{align}
\log p(o) = \mathcal{D}[q(s)||p(s|o)]  - \mathcal{F}[q(s)],
\end{align}
where $\mathcal{F}[q(s)]$ is the \emph{variational free energy}:
\begin{align}
\mathcal{F}[q(s)] = \sum_{s} q(s) \log \frac{q(s)}{p(o,s)}.
\end{align}
The free energy is equivalent to the negative of the \emph{evidence lower bound}, the more common term in the machine learning literature \cite{blei17}.

Note that the free energy only requires knowledge of $p(s|o)$ up to a normalizing constant, since $p(s|o) \propto p(o|s) p(s)$. This is typically unproblematic, since we can often compute the prior $p(s)$ and likelihood $p(o|s)$ of any particular state $s$. Critically, the identity above implies that minimizing the free energy is equivalent to minimizing KL divergence, since the two must balance each other out to match the marginal likelihood, which is fixed as a function of $q$. Thus, minimizing free energy when the variational family is unrestricted is equivalent to exact Bayesian inference.

If FEP = Bayes, then we cannot distinguish its predictions from other asymptotically correct inference algorithms, such as Monte Carlo sampling, except when these algorithms are restricted in some way. Monte Carlo methods may, for example, be restricted in terms of the number of samples they generate or how they generate the samples (e.g., \cite{dasgupta17}). Optimization of free energy is typically restricted by placing constraints on the variational family, as we  discuss next.

\section{Restricting the variational family}

If the hypothesis space is vast, then summing (or integrating) over all possible hypotheses to compute the free energy will be intractable. Thus, essentially all practical applications of free energy optimization make use of a restriction on $\mathcal{Q}$ that renders the optimization tractable (as will be discussed below). The important point for present purposes is that as long as the true posterior is in $\mathcal{Q}$, the optimal $q^\ast$ will be equal to the posterior. Thus, FEP in its most general form is indistinguishable from Bayesian inference.

Practical applications of free energy optimization restrict $\mathcal{Q}$ in some way to make the problem tractable. These restrictions typically mean that the posterior is no longer contained in $\mathcal{Q}$, and thus the distribution that minimizes free energy will deviate from Bayes-optimality: $q^\ast(s) \neq p(s|o)$.

The widely used ``mean-field'' approximation assumes that the posterior factorizes across components of $s$ (i.e., dimensions of the state space):
\begin{align}
q(s) = \prod_{i} q_i(s_i).
\end{align}
For example, if I'm trying to infer the posterior over the height and weight of an individual given their gender, I could assume that the posterior factorizes into $q(\text{height}|\text{gender})$ and $q(\text{weight}|\text{gender})$. Because the true posterior rarely factorizes, the mean-field approximation will produce systematic errors. For example, if the factorization is across a sequence of states, the posterior may be biased by the order of the data. Intriguingly, these errors can be discerned in human behavior \cite{daw08,sanborn13}. On the other hand, the mean-field approximation may work well in many cases, which is why it is widely adopted in machine learning. This effectiveness can render it difficult to test as a process model, because it often makes similar predictions to exact Bayesian inference.

When $s$ is continuous, another common restriction is to assume that the posterior is Gaussian \cite{friston07}, parametrized by a mean $\mu$ and covariance matrix $\Sigma$:
\begin{align}
q(s) = \mathcal{N}(s; \mu,\Sigma).
\end{align}
These parameters are then chosen to minimize the free energy, typically by gradient descent. The Gaussian approximation can be motivated by the ``Bayesian central limit theorem,'' which states that the posterior is approximately Gaussian around the mode when the amount of data is large relative to the dimensionality of $s$. It can also be generalized to mixtures of Gaussians to approximate multimodal posteriors \cite{gershman12}.

One challenge facing applications of the Gaussian approximation is that the free energy is not, in general, tractable (except in the case where the exact posterior is Gaussian). To deal with this issue, a common technique, known as the \emph{Laplace approximation}, is to use a second-order Taylor series expansion around the posterior mode. This replaces the nonlinear free energy with a quadratic function, rendering the free energy tractable. The price we pay for this approximation is that we are no longer optimizing the free energy, and we have no guarantee that this will produce sensible answers, or even converge. It turns out, however, that the Laplace approximation has intriguing implications for the neurobiological implementation of Bayesian inference.

\section{Predictive coding}

The Laplace approximation can be used to derive arguably the most influential and distinctive aspect of FEP---\emph{predictive coding}, according to which feedback pathways convey predictions, and feedforward pathways in the brain convey prediction errors (discrepancies between data and predictions). The idea of predictive coding has a long history in signal processing \cite{elias55}, and was previously proposed as a theory of redundancy reduction (efficient coding) in neural signals \cite{rao99}. Friston and colleagues showed how predictive coding could be derived within the framework of free energy minimization \cite{friston07,friston08,friston09}, how it could be mapped onto the structure of biologically realistic microcircuits \cite{bastos12}, and how it could be applied to motor control \cite{friston11} and action selection more generally (a topic we visit in the next section).

Friston and colleagues started from the following assumptions:
\begin{itemize}
\item The internal (generative) model is hierarchically structured, such that hidden states at higher levels generate hidden states at lower levels.
\item The approximate posterior factorizes across hidden state dimensions within and between levels of the internal model (i.e., the mean-field approximation).
\item Each component of the factorized posterior is modeled as a Gaussian.
\end{itemize}
They then used the Laplace approximation to approximate the free energy and derive update rules for optimization based on gradient descent. They showed that this optimization scheme corresponds to a form of predictive coding, which is found ubiquitously in the engineering literature (e.g., Kalman filtering).

It is important to emphasize that predictive coding is \emph{not} a generic consequence of FEP, or even of FEP with a specific approximation family. It is derived from a combination of assumptions about the internal model (hierarchical organization), the approximation family (factorized and Gaussian), the approximation of the free energy (quadratic around the mode), and the optimization scheme (gradient descent). With all of these assumptions in place, FEP does make claims that go beyond the general Bayesian brain hypothesis, and have received ample empirical support \cite{aitchison17,murray02,summerfield08,egner10,kok15}. Alternatively, some authors have explored variants of FEP that do not invoke predictive coding, or combine it with other neural message passing schemes (e.g., \cite{friston17}).

\section{Active inference}

Let us return now to the setting in which an agent can take actions (according to policy $\pi$) to influence its observations. In this setting, FEP posits that the agent seeks to minimize \emph{expected} free energy under future observations $o$ and future state $s$ \cite{friston15}:
\begin{align}
\sum_{o} p(o|s,\pi) \sum_s q(s|\pi) \log \frac{q(s|\pi)}{p(o,s|\pi)} = - \sum_o q(o|\pi) \mathcal{D}[q(s|o,\pi)||q(s|\pi)]  - \sum_o q(o|\pi) \log p(o|\pi),
\label{eq:EFE}
\end{align}
where $q(s|\pi)$ is the approximate belief about future state $s$ prior to observing $o$. Friston and colleagues refer to the minimization of expected free energy with respect to actions as \emph{active inference}. Note that here the likelihood is stipulated to be $p(o|s,\pi) = q(o|s,\pi)$, and we have assumed that the predictive posterior $q(s|o,\pi) \approx p(s|o,\pi)$.

When the approximate posterior is exact, the first term in the expression is the negative information gain and the second term is the entropy $\mathcal{H}[p(o|\pi)]$ of the future observations conditional on the policy:
\begin{align}
\sum_{o} p(o|s,\pi) \sum_s p(s|\pi) \log \frac{p(s|\pi)}{p(o,s|\pi)}  = -\mathcal{I}(\pi)  + \mathcal{H}[p(o|\pi)].
\label{eq:EFE2}
\end{align}
If in addition observations are deterministic functions of the policy, then the entropy term is 0, and minimizing expected free energy is equivalent to maximizing information gain. Thus, under certain conditions active inference is equivalent to the information gain policy studied in standard Bayesian treatments of information acquisition \cite{nelson05}. When the observations are stochastic and can be interpreted as reward outcomes (see next section), active inference instantiates a form of risk-sensitive control, since actions that reduce outcome variability will be favored (see \cite{friston15} for more discussion). Another way of thinking about the entropy term is that it reflects the ``coding cost'' of unpredictable data, since entropy is a lower bound on the average number of bits needed to communicate observations via a sensory channel without loss of information \cite{shannon48}. Thus, active inference prefers actions that produce observations which are both informative and predictable.

As in the previous sections, we can ask which aspects of this analysis are generic implications of the Bayesian brain hypothesis (with an information gain policy), and which are specific to FEP. We showed that FEP is equivalent to Bayesian information gain only under the special case of an exact posterior and deterministic outcomes in the future. When the determinism constraint is relaxed, information gain and expected free energy will be substantively different.

\section{Planning as inference}

A number of papers on active inference make an additional conceptual move (e.g., \cite{friston09b,friston12,friston15}), reinterpreting the entropy term as a form of \emph{extrinsic value}, contrasting it with the \emph{epistemic value} of the information gain term. Central to this reinterpretation is the postulate that the utility of an outcome is equal to its log prior probability, $u(o) = \log p(o|\pi)$, usually referred to as its \emph{prior preference}. (Note that we are conditioning on the policy here to emphasize that the free energy is being computed for a fixed policy.) This leads to a form of \emph{planning as inference} \cite{botvinick12,kappen12}, whereby minimizing free energy optimizes a combination of expected utility (extrinsic value) and information gain (epistemic value).

At first glance, this seems rather odd; why should utility be proportional to probability? Undoubtedly there are high probability events that have low utility (e.g., if you are born into poverty then lacking access to basic goods may be highly probable). However, note that this is potentially just Bayesian decision theory in disguise: as long as I'm allowed to choose probabilities that are proportional to utilities, FEP will coincide with Bayesian decision theory. The critical step in this logic is the assumption that evolution has equipped us with the belief that low utility states are low probability, due to the fact that if our ancestors spent a lot of time in those states they would be less likely to reproduce. Whether or not this is a reasonable assumption, the technical point is that planning as inference can be understood as a notational variant of Bayesian decision theory, provided the utilities and probabilities coincide (free energy theorists typically stipulate that they coincide). FEP can make distinctive predictions when they don't coincide, or when the planning as inference transformation leads to different algorithmic approximations or neural implementations, provided the utilities and prior preferences coincide (i.e., effectively, replacing utility with prior preferences).

\section{Conclusions}

There are several take-home messages from this article:
\begin{itemize}
\item For passive observations (no actions), the predictions of FEP are indistinguishable from the predictions of the Bayesian brain hypothesis when the variational family is unrestricted (i.e., the when the exact posterior is in the variational family, and hence minimizing free energy is equivalent to exact inference).
\item Predictive coding is not a generic consequence of FEP; it arises only under certain restrictions of the variational family and a specific choice of optimization scheme.
\item In the active setting (observations can be influenced by actions), active inference is equivalent to an information gain policy when the approximate posterior is exact and the observations are deterministic functions of actions. When observations are stochastic, active inference induces a form of risk-aversion not found in the information gain policy.
\item When utilities are interpreted as log probabilities, FEP corresponds to a form of planning as inference, a class of algorithms for utility maximization. The predictions of FEP are distinguished from utility maximization when utilities don't correspond to log probabilities.\footnote{Technically, utilities can always be formulated as log probabilities. But it is an empirical question whether subjective beliefs and preferences disclosed by behavior coincide \cite{gershman12b}.}
\item When utilities are interpreted as prior preferences, FEP places value on information gain. This also arises naturally in Bayesian decision theory applied to sequential decision problems and hence is not a distinctive prediction.
\end{itemize}

These take-home messages do not exhaust the set of ideas that have been introduced under the banner of FEP. For example, FEP has been offered as a first-principle account of self-organization \cite{friston13} and ecological niche construction \cite{constant18}. We have focused here on issues that are more central to neuroscience.

The broader point of this article is that a unifying theory like FEP needs to be deconstructed in order to be properly evaluated and compared to alternative theories. By undertaking part of this deconstruction, we hope to make the elegant synthesis offered by FEP more accessible to the broader neuroscience community.

\subsection*{Acknowledgments}

I am grateful to Ben Vincent, Momchil Tomov, Chris Summerfield, Giovanni Pezzulo, Peter Battaglia, Jan Drugowitsch, Rani Moran, Yuqing Hou, Jascha Achterberg, Robert Rosenbaum, Sabya Shivkumar, and Nathaniel Daw for comments on an earlier draft of the paper.

\bibliographystyle{rss}

\end{document}